\documentclass[10pt,twocolumn,letterpaper]{article}

\usepackage{cvpr}              

\usepackage[dvipsnames]{xcolor}

\usepackage{caption}
\usepackage[accsupp]{axessibility} 

\definecolor{cvprblue}{rgb}{0.21,0.49,0.74}
\usepackage[pagebackref,breaklinks,colorlinks,citecolor=cvprblue]{hyperref}

\def\paperID{5346} 
\def\confName{CVPR}
\def\confYear{2024}

\title{Selective nonlinearities removal from digital signals}

\author{Krzysztof A. Maliszewski\\
School of Mathematics and Statistics, University of Canterbury\\
Christchurch, New Zealand\\
{\tt\small k.a.maliszewski@gmail.com}
\and
Magdalena A. Urbańska\\
School of Food and Advanced Technology, Massey University\\
Palmerston North, New Zealand\\
{\tt\small murbanska@massey.ac.nz}
\and
Varvara Vetrova\\
School of Mathematics and Statistics, University of Canterbury\\
Christchurch, New Zealand\\
{\tt\small varvara.vetrova@canterbury.ac.nz}
\and
Sylwia M. Kolenderska\\
Institute of Physics, Faculty of Physics, Astronomy and Informatics\\
Nicolaus Copernicus University in Torun\\
ul. Grudziadzka 5, 87-100 Torun, Poland\\
School of Physical and Chemical Sciences, University of Canterbury\\
Christchurch, New Zealand\\
{\tt\small skol745@aucklanduni.ac.nz}
}

\begin{document}
\maketitle
\begin{abstract}
Many instruments performing optical and non-optical imaging and sensing, such as Optical Coherence Tomography (OCT), Magnetic Resonance Imaging or Fourier-transform spectrometry, produce digital signals containing modulations, sine-like components, which only after Fourier transformation give information about the structure or characteristics of the investigated object. Due to the fundamental physics-related limitations of such methods, the distribution of these signal components is often nonlinear and, when not properly compensated, leads to the resolution, precision or quality drop in the final image. Here, we propose an innovative approach that has the potential to allow cleaning of the signal from the nonlinearities but most of all, it now allows to switch the given order off, leaving all others intact. The latter provides a tool for more in-depth analysis of the nonlinearity-inducing properties of the investigated object, which can lead to applications in early disease detection or more sensitive sensing of chemical compounds. We consider OCT signals and nonlinearities up to the third order. In our approach, we propose two neural networks: one to remove solely the second-order nonlinearity and the other for removing solely the third-order nonlinearity. The input of the networks is a novel two-dimensional data structure with all the information needed for the network to infer a nonlinearity-free signal. We describe the developed networks and present the results for second-order and third-order nonlinearity removal in OCT data representing the images of various objects: a mirror, glass, and fruits.
\end{abstract}    
\section{Introduction}
\label{sec:intro}

\begin{figure}[t]
  \centering
  \includegraphics[width=0.85\linewidth]{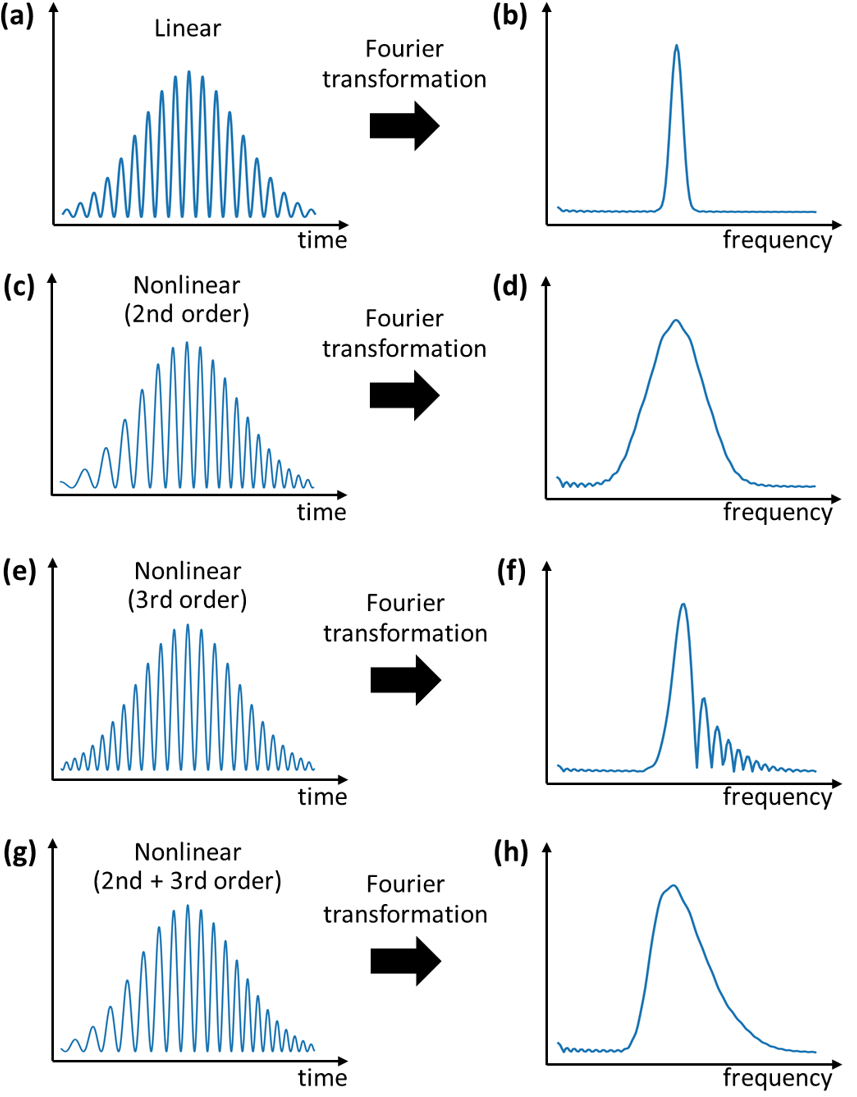}
  \captionsetup{belowskip=-3.5mm}
  \caption{Effects of signal nonlinearity on the signal's Fourier transform.
 (a) A signal with a linear modulation distribution Fourier transform to (b) a narrow peak. (c) If the nonlinearity of the modulation is of the second order, (d) the peak gets broadened. (e) If the nonlinearity is of the third order, (f) the peak gets distorted. In OCT, this will result in deteriorated image quality. (g) In practice, both nonlinearities appear at once, resulting in (h) the peak being elongated in the direction of the third-order distortion.}
  \label{fig:nonlin-effects}
\end{figure}

Optical Coherence Tomography (OCT), Magnetic Resonance Imaging (MRI) or Fourier-transform Spectrometry (FTS) belong to a class of methods where useful, interpretable signal is obtained in an indirect way, more specifically by Fourier transforming the raw signal acquired by the instrument and taking its absolute value. On the one hand, such an indirect approach makes OCT-based imaging and FTS-based sensing faster and %
enables MRI, but on the other hand, it makes room for errors related to the limitations of Fourier transformation. 
One such limitation is the Fourier transformation's sensitivity to nonlinearities in the input signal. In the simplest case of a signal being a single-frequency modulation, a second-order nonlinear distribution 
will Fourier transform to a broader peak (compare \cref{fig:nonlin-effects}a,b with c,d). If this distribution is of a third-order nonlinear character, the peak in the Fourier transform will get distorted (\cref{fig:nonlin-effects}e,f). In practice, both nonlinearities appear at once, the second-order one bigger than the third-order one, leading to a broad peak elongated in the direction of the third-order nonlinearity distortion (\cref{fig:nonlin-effects}g,h).

The influence of nonlinearities is best visible in OCT, where the Fourier transform of the raw signal constitutes one line of the image. Since the width of the peak represents the depth resolution (i.e. how small a detail can be observed), and the peak's shape correlates with the image quality, nonlinearities are generally thought of as detrimental and various techniques are devised to remove them \cite{wang2008spectral,lan2017design}. These techniques do not achieve absolute removal of nonlinearity,  preventing single-shot imaging of bulk objects, such as the entire eye \cite{grulkowski2018swept,hillman2005effect} and consequently necessitating elaborate workaround solutions \cite{chen2022high,huang2023physical,marks2003digital,pan2017depth}. A Machine Learning solution for nonlinearity removal was proposed as well \cite{ahmed2022adc} – although robust and reliable, it is biased towards specific kinds of objects with specific nonlinear properties. 

On the other hand, the inherent optical properties that produce nonlinearities in the imaged objects have been shown to enable the characterisation of the objects \cite{kolenderska2018dispersion,photiou2019comparison,maliszewski2023extracting} and correlate the change of their nonlinear character with the progression of diseases \cite{photiou2017using}. Unfortunately, the current methods for extracting the second-order nonlinearity are very object-specific as well as error-prone preventing them from being applied in real-world scenarios.

We present an approach that enables flexible and comfortable management of the signal nonlinearities, letting the user study one order in the absence of the others, a capability previously impossible to achieve using current standard or Machine Learning methods. In our approach, for each nonlinearity order, a neural network is trained whose input is a special two-dimensional data structure, a stack. Each row of the stack represents an amplitude of the Fourier transform of an input signal multiplied by a different nonlinearity-correcting factor. Consequently, it contains nonlinearity-free regions which the network identifies and uses to build a nonlinearity-free Fourier transform amplitude. We present two networks, one to remove the second-order nonlinearity while keeping the third-order one intact, and the second network to remove the third-order nonlinearity while keeping the second-order one intact. We test our approach on OCT signals that naturally contain substantial amounts of second-order nonlinearity and small amounts of third-order nonlinearity. The tests on both computer-generated and experimental signals indicate the robustness of our approach.

We start by describing our approach in Section \ref{sec:concept}, including the creation of the stacks in Subsection~\ref{subsec:approach-1}, their application in the training processes in Subsection~\ref{subsec:nonlinearity_removing_networks},  presentation of the used architecture and methodology in Subsection~\ref{subsec:methodology}, and the description of the datasets (Subsection~\ref{subsec:datasets}) used for training of our models. In Section~\ref{sec:performance}, we report the performance of the networks in removing the respective nonlinearity, complete with the analysis of the training process and the subsequent tests on computer-generated data (Subsection~\ref{subsec:in-sillico}) and experimental data (Subsections~\ref{subsec:mirrors}, \ref{subsec:glass} and \ref{subsec:grapecucumber}). In the final sections, we summarise the obtained results (Section~\ref{sec:summary}), pointing to the strong points of our approach, identifying the observed and potential shortcomings, and discussing potential solutions and the scope of future work (Section~\ref{sec:future}).
\section{Concept} \label{sec:concept}

\subsection{Amplitude stacks as inputs} \label{subsec:approach-1}

\begin{figure}[t]
    \centering
    \includegraphics[width=1.0\linewidth]{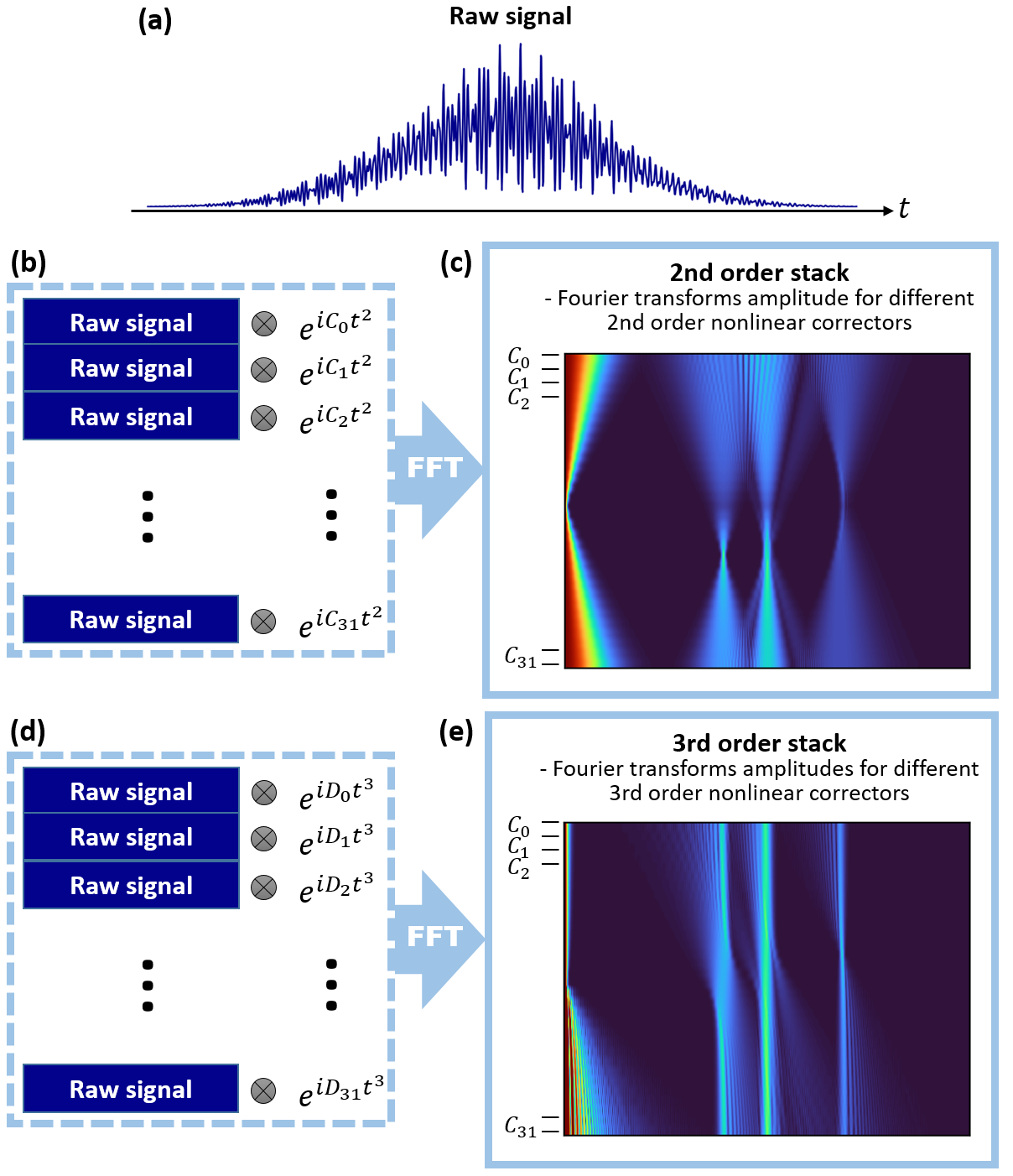}
    \captionsetup{belowskip=-3mm}
    \caption{Calculation of the input stacks.
  (a) The raw signal is (b) multiplied by 32 different complex exponents representing 32 different levels of the second-order nonlinearity: $C_0$, $C_1$, $C_2$, ..., $C_{31}$. Each row is then Fourier transformed. (c) The 2nd-order stack is created from the absolute value of the complex Fourier transforms. Similarly, (d) the raw signal is multiplied by 32 different complex exponents representing 32 different levels of the third-order nonlinearity: $D_0$, $D_1$, $D_2$, ..., $D_{31}$ and then Fourier transformed to build (e) a 3rd-order stack.}
    \label{fig:stacks}
\end{figure}

\begin{figure}[t]
    \centering
    \includegraphics[width=1.0\linewidth]{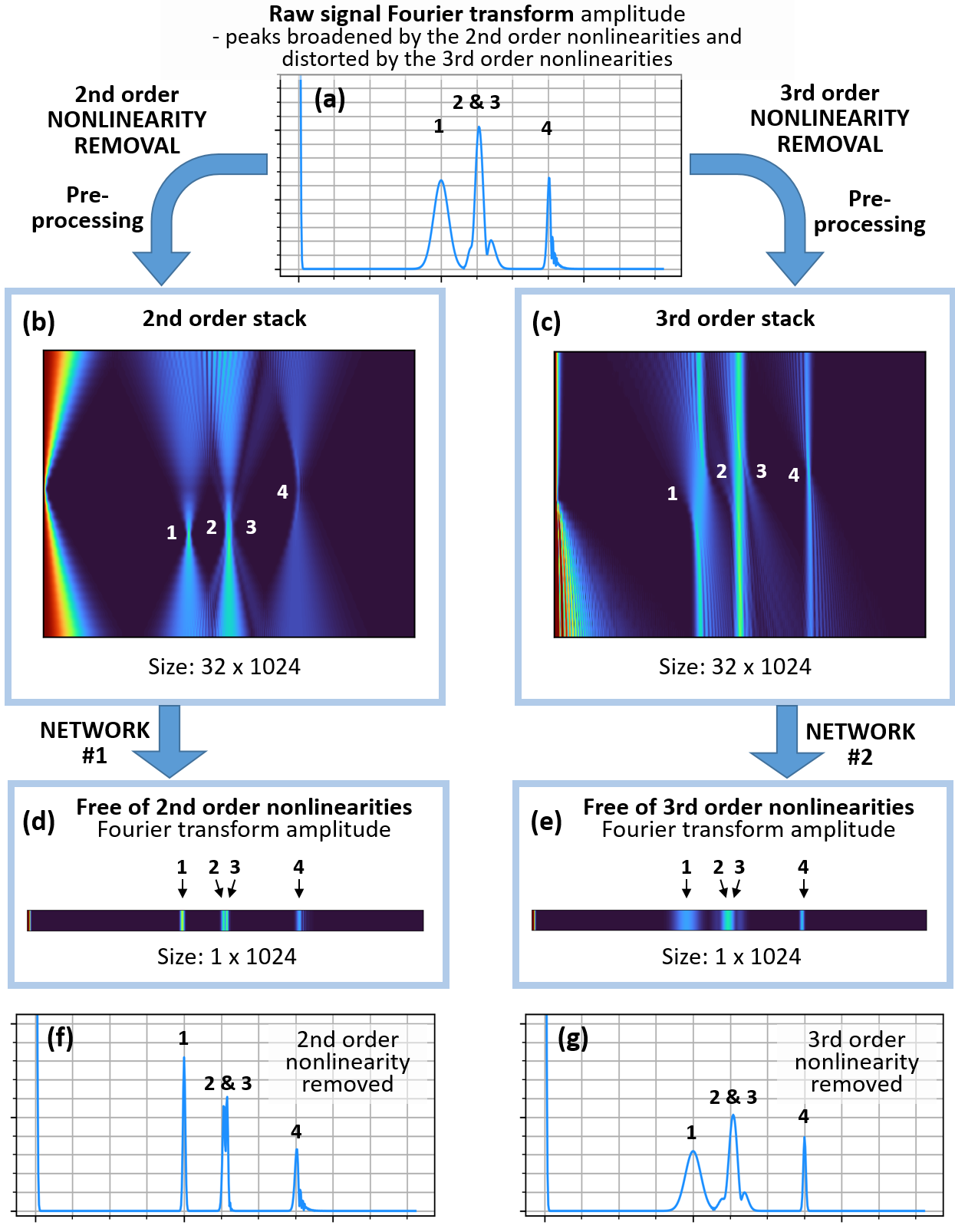}
    \captionsetup{belowskip=-3mm}
    \caption{
  (a) The Fourier transform amplitude of a raw signal, presented in \cref{fig:stacks}a, shows four peaks affected by a different amount and type of nonlinearity. The second-order nonlinearity is removed with Network~\textbf{\#1} 
  trained on 
  (b) the 2nd-order stack as input and (d) Fourier transform amplitude free of the second-order nonlinearity as output. Similarly, the third-order nonlinearity is removed with Network~\textbf{\#2} 
  trained on (b) the 3rd-order stack as input and (d) Fourier transform amplitude free of the third-order nonlinearity as output. As a result, (f) the second-order nonlinearity is removed leaving the third-order one (see peak 4) using Network~\textbf{\#1}, (g) the third-order nonlinearity is removed leaving the second-order one intact (see peak 4) using Network~\textbf{\#2}.}
    \label{fig:networks}
\end{figure}

Our networks rely on a two-dimensional data structure as inputs: a stack (\cref{fig:stacks}). The rows of a stack are created in the following way. First, the raw signal is multiplied (\cref{fig:stacks}a) by a complex exponent with a different quadratic argument in the case of the stack used for removing the second-order nonlinearity (\cref{fig:stacks}b) or with a different cubic argument in the case of the stack for the third-order nonlinearity removal (\cref{fig:stacks}d). Coefficients in the quadratic arguments, ($C_{0},...,C_{31}$ in \cref{fig:stacks}b), and coefficients in the cubic arguments ($D_{0},...,D_{31}$ in \cref{fig:stacks}d), are chosen to cover a wide range of the second- and thrid-order nonlinearity, ensuring that the nonlinearities present in the signal are fully compensated. Then, each row is Fourier transformed, and an absolute value is calculated to obtain a 2nd-order stack (\cref{fig:stacks}c) or a 3rd-order stack (\cref{fig:stacks}e). The range of the nonlinear coefficients is symmetrical around 0 and broad enough to ensure that each modulation's nonlinearity in the raw signal is compensated. In the 2nd-order stack, these will be the places with the smallest width (marked 1, 2, 3 and 4 in \cref{fig:networks}b) and in the 3rd-order stack, the places containing no distortions (marked 1, 2, 3 and 4 in \cref{fig:networks}c).

\subsection{Nonlinearity-removing networks} \label{subsec:nonlinearity_removing_networks}

The approach for removing the second-order nonlinearity and the approach for removing the third-order nonlinearity are similar, presented schematically in \cref{fig:networks}. The input of the network for the second-order nonlinearity removal, Network~\textbf{\#1}, is the 2nd-order stack (\cref{fig:networks}b) and the output is a Fourier transform amplitude free of second-order nonlinearity (\cref{fig:networks}d). Because the input 2nd-order stack contains regions corresponding to perfectly removed second-order nonlinearity, the training process can be viewed as learning how to identify these regions and then stitching them all up. Similarly, the network for removing third-order nonlinearity, Network~\textbf{\#2}, takes the 3rd-order stack as input (\cref{fig:networks}c) and outputs a Fourier transform amplitude free from the 3rd-order nonlinearity (\cref{fig:networks}e). Input for both networks has a shape of 32 by 1024 pixels with a single channel whereas output is a vector of size 1024. When compared to the Fourier transform amplitude of the raw signal (\cref{fig:networks}a), the signal returned by Network~\textbf{\#1} (\cref{fig:networks}f) incorporates only the third-order nonlinearity, and the signal returned by Network~\textbf{\#2} (\cref{fig:networks}g) contains only the second-order nonlinearity. 


\subsection{Methodology} \label{subsec:methodology}

Our nonlinearity-removing networks are inspired by the U-Net architecture \cite{ronneberger2015u}. U-Net stands out for its use of a contracting path to capture contextual information, an expanding path for precise feature localisation, and the incorporation of skip connections that preserve fine-grained details, particularly beneficial for precise object delineation. We leverage these capabilities to identify and extract nonlinearity-related features from the stacks.

\begin{figure}[t]
    \centering
    \includegraphics[width=1.0\linewidth]{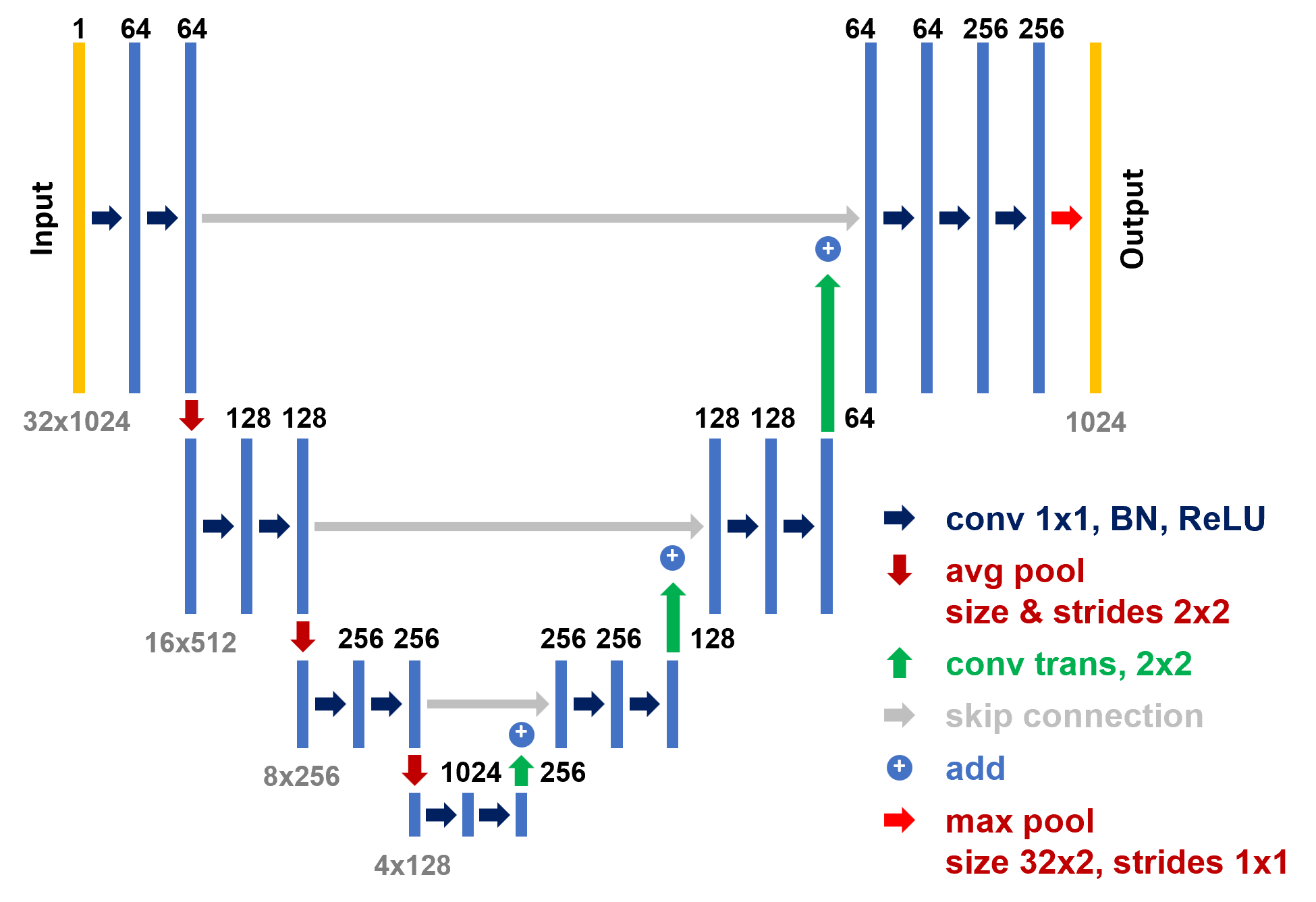}
    \captionsetup{belowskip=-3mm}
    \caption{Architecture of our nonlinearity-removing networks.}
    \label{fig:architectures}
\end{figure}

While the classical U-Net deals with 2D data structures for both input and output, in our case, we perform dimension reduction to output 1D signals by incorporating a max-pooling layer. We further refined the architecture by optimising hyperparameters with the validation dataset using Optuna \cite{akiba2019optuna}. 
As a result of the tuning, we adopted the Mean Absolute Error (MAE) loss function and the Adam optimiser \cite{kingma2014adam} with a set learning rate of 0.0002. Our final network architecture is presented in \cref{fig:architectures}. 

\subsection{Datasets} \label{subsec:datasets}

Our datasets consist of noiseless computer-generated signals that are Gaussian-shaped, and diverse and randomised in terms of their content. If put in the OCT-imaging-related terms, they represent a broad range of possible imaged object configurations. These objects are simulated to have a random number of interfaces (from 2 to 12) positioned at arbitrary locations, with each object layer exhibiting varying random reflectivity and second- and third-order nonlinearity values. These object parameters are used to synthesise modulatory signals equivalent to those acquired in an OCT measurement (see example signals in \cref{fig:nonlin-effects}a, c, e, g or \cref{fig:stacks}a). Next, the input stacks and outputs, the Fourier transform amplitudes of signals free of the respective nonlinearity, are calculated.

The risk of overfitting was minimised by using the training datasets with an optimised size \cite{barbedo2018impact}. In our research, this was found to be between 160,000 and 240,000 objects. The validation and test datasets include 14,993 objects. The distribution is uniform across the objects with a specific number of interfaces, resulting in an equal allocation of 1,363 data samples for each distinct object type.

Before starting the training, as part of the preprocessing stage, both the input and output data were normalised through Min-Max scaling, ensuring that values fell within the standardised range of 0 to 1.


\section{Performance} \label{sec:performance}

\subsection{In silico data tests} \label{subsec:in-sillico}

\begin{figure}[t]
    \centering
    \includegraphics[width=1.0\linewidth]{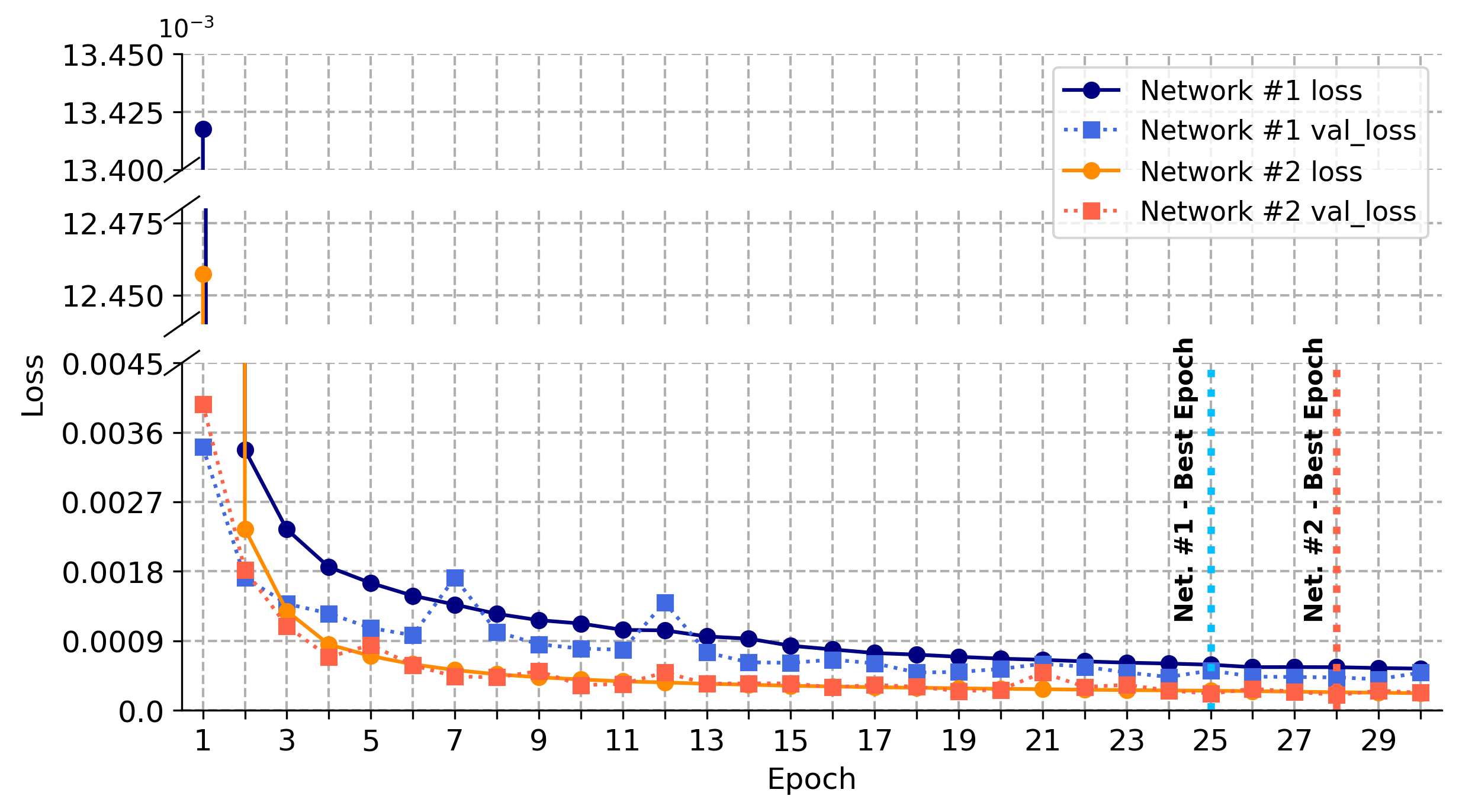}
    \vspace{-3mm}
    \captionsetup{belowskip=-3mm}
    \caption{ Network~\textbf{\#1} and Network~\textbf{\#2} performance measured with the Mean Absolute Error (MAE) loss function across 30 epochs of training. Blue lines depict Network~\textbf{\#1}, while orange lines showcase the Network~\textbf{\#2} performance. Dotted vertical lines highlight epochs exhibiting the best performance.}
    \label{fig:loss}
\end{figure}

Each nonlinearity-removing neural network underwent training using a training dataset consisting of 200,000 objects. Within each mini-batch, 8 input stacks were processed. The proposed algorithm requires 39 ms to process a stack on a Quadro RTX 6000 GPU. Network~\textbf{\#1} (\cref{fig:loss}, blue lines) and Network~\textbf{\#2} (\cref{fig:loss}, orange lines) both displayed convergence within just a few epochs (\cref{fig:loss}) and were trained until the thirtieth epoch. 

To better evaluate the networks and results, we employed the Goodness-of-Fit Approximity (GoF) metric introduced in \cite{maliszewski2023extracting}. The GoF metric assesses the accuracy of predictions by calculating the percentage of values that lie within a predefined distance from the corresponding ground truth values. For this GoF metric, we define an error threshold as a ratio between the predefined distance and the value range span. In our case, where the values range from 0 to 1, the distance of 0.01 means 1\% error threshold.

Based on the post-training evaluation using a 1\% error threshold and the test dataset, Network~\textbf{\#1} started to overfit at epoch 25 (\cref{fig:loss}, indicated by the blue dotted vertical line). In contrast, Network~\textbf{\#2} demonstrated resilience against overfitting and achieved its best results in the 28th epoch (\cref{fig:loss}, marked by the red dotted vertical line). 

\begin{figure}[t]
    \centering
    \includegraphics[width=1.0\linewidth]{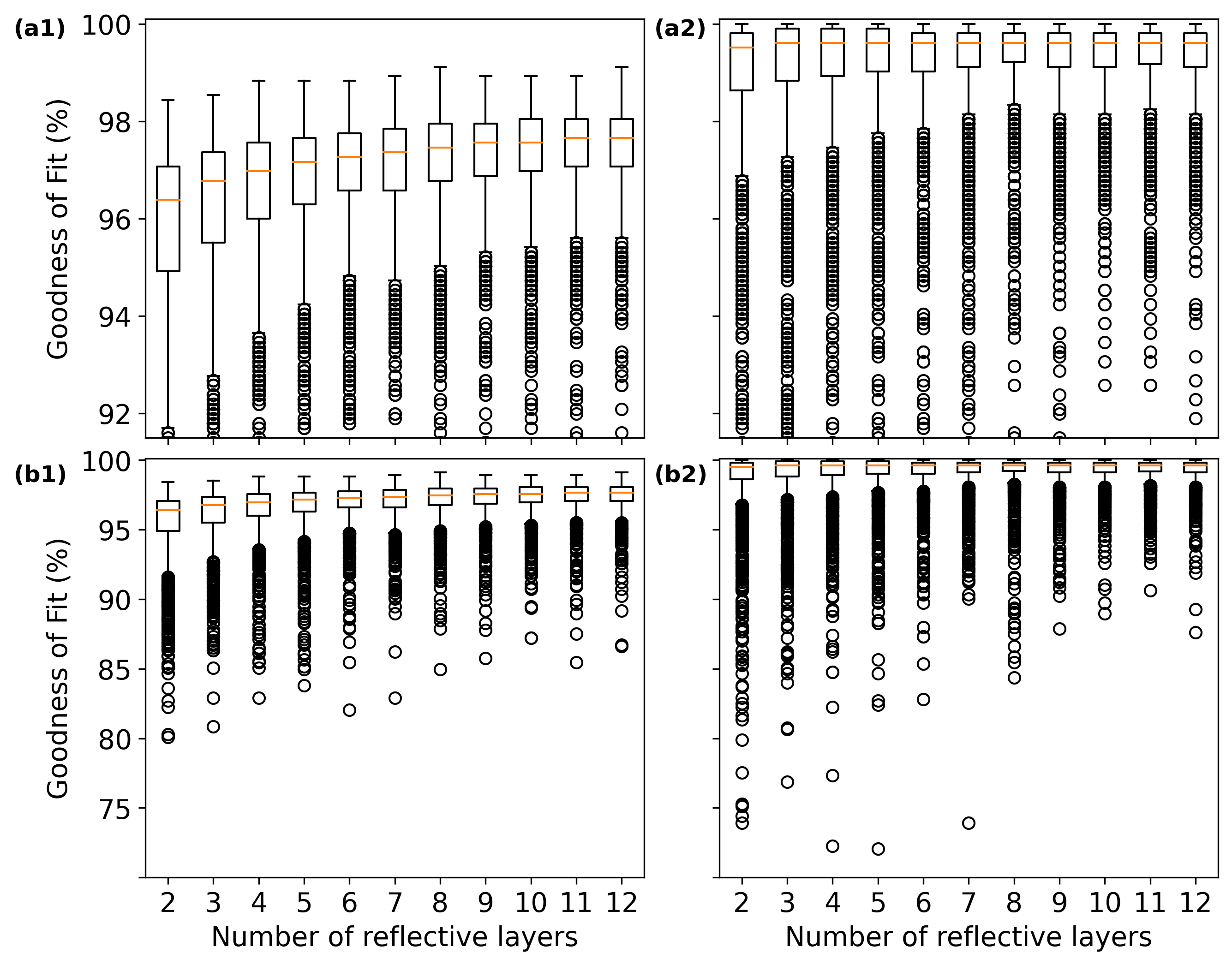}
    \captionsetup{belowskip=-3mm}
    \caption{Networks performance assessed through the metric of GoF, measuring the degree to which predictions align with the ground truth within 0.1\% error in the test dataset comprising 14,993 data samples for (a) Network~\textbf{\#1} and (b) Network~\textbf{\#2}. The graphs at the top (a1,a2) are zoomed-in areas of the top area of (b1,b2) the whole graphs presented.}
    \label{fig:gof}
\end{figure}

Using the test dataset and GoF with a 1\% error threshold, both networks achieved a remarkable mean GoF value of 99.98\%, which means that 99.98\% of the predictions differed from the ground truth by less than 1\%, with only one example below 95\% GoF. In light of this, to be able to carry out a meaningful analysis, the error threshold was further reduced to 0.1\% and a boxplot graph was plotted for each model in \cref{fig:gof}. 

\begin{figure}
    \centering
    \includegraphics[width=1.0\linewidth]{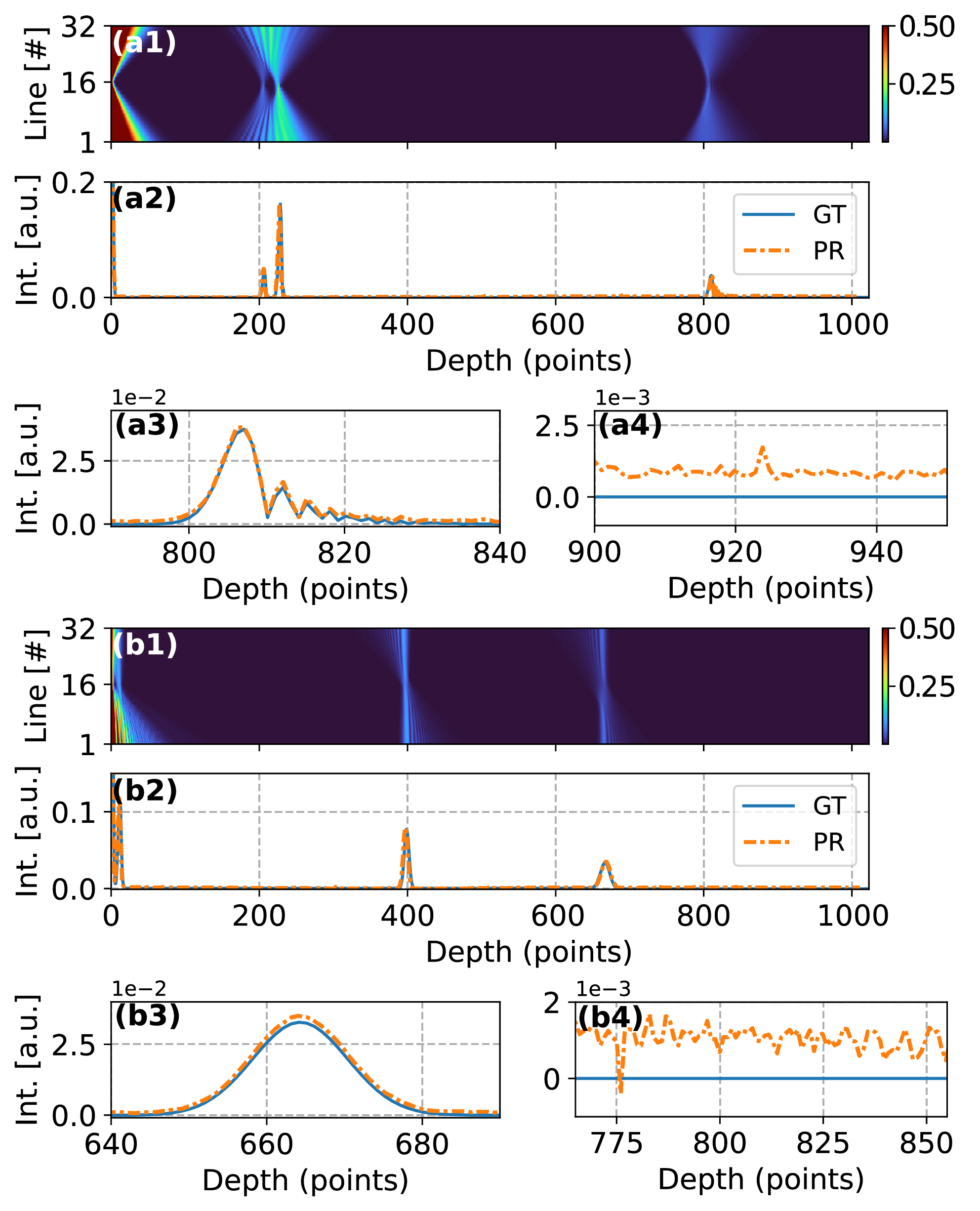}
    \caption{Predictions with very low GoF for 0.1\% error threshold for (a) Network~\textbf{\#1} and (b) Network~\textbf{\#2}. Based on the input stack (1), models predict (2, orange line, PR) amplitudes free of the respective nonlinearity that exhibit a strong correlation with the ground truth (2, blue line, GT). (3) Nonlinearity-free peaks are recovered with high fidelity, but (4) the models sometimes struggle with zero-valued places.}
    \label{fig:comparison}
\end{figure}

Both networks, Network~\textbf{\#1} (\cref{fig:gof}a) and Network~\textbf{\#2} (\cref{fig:gof}b), demonstrated very good performance even for error threshold 0.1\%. Only approximately  9.52\% and 3.07\% of the predictions from Network~\textbf{\#1} and Network~\textbf{\#2}, respectively, fell below the 95\% GoF for such a small error, as showed in \cref{tab:below_gof95} that also includes GoF values for objects with distinct number of layers. \cref{fig:comparison} offers an in-depth analysis of the worst-predicted examples for the Network~\textbf{\#1} (\cref{fig:comparison}a) and the Network~\textbf{\#2} (\cref{fig:comparison}b) showcasing GoF values of 65.5\% and 45.5\%, respectively. Predictions of peak amplitudes in the nonlinearity-free Fourier transform amplitude (\cref{fig:comparison}(3), orange line) closely align the expected outcomes (blue line) deviating only by the error threshold level. Also, the networks predict values close to zero rather than precisely zero, contributing to worse GoF metric results. One should keep in mind here that these discrepancies are extremely low: at the level of around 2e-3 as presented in \cref{fig:comparison}a4 and b4.

\begin{table*}
    \caption{The number of cases where GoF with error threshold 0.1\% falls below 95\% for objects with a specific number of reflective layers.}
    \label{tab:below_gof95}
    \centering
    \scalebox{0.8}{
        \begin{tabular}{llllllllllllll}
Net. & Interfaces & 2 & 3 & 4 & 5 & 6 & 7 & 8 & 9 & 10 & 11 & 12 & Totals\\
\hline\hline
 & \# below 95\% &  384 &    260 &    193 &    138 &    105 &   87 &    79 &    55 & 45 & 44 & 37 & 1381 \\
\textbf{\#1} & \# in dataset  &  1363 &    1363 &    1363 &    1363 &    1363 &   1363 &    1363 &    1363 & 1363 & 1363 & 1363 & 14993 \\
& \% below & 28.17 & 19.08 & 14.16 & 10.12 & 7.70 & 6.38 & 5.80 & 4.04 & 3.30 & 3.32 & 2.71 & \textbf{9.52} \\
\hline\hline
 & \# below  95\%  &  107 &  82 &  57 &  54 &  38 &  32 &  31 &  22 & 13 & 13 & 11 & 460 \\
\textbf{\#2} & \# in dataset  &  1363 &    1363 &    1363 &    1363 &    1363 &   1363 &    1363 &    1363 & 1363 & 1363 & 1363 & 14993 \\
& \% below & 7.85 & 6.02 & 4.18 & 3.96 & 2.79 & 2.35 & 2.27 & 1.61 & 0.95 & 0.95 & 0.81 & \textbf{3.07} \\
\hline
\end{tabular}
}
\vspace{-4mm}
\end{table*}

\subsection{Experimental signals: mirror at different depths} \label{subsec:mirrors}

\begin{figure}[t]
    \centering
    \includegraphics[width=0.85\linewidth]{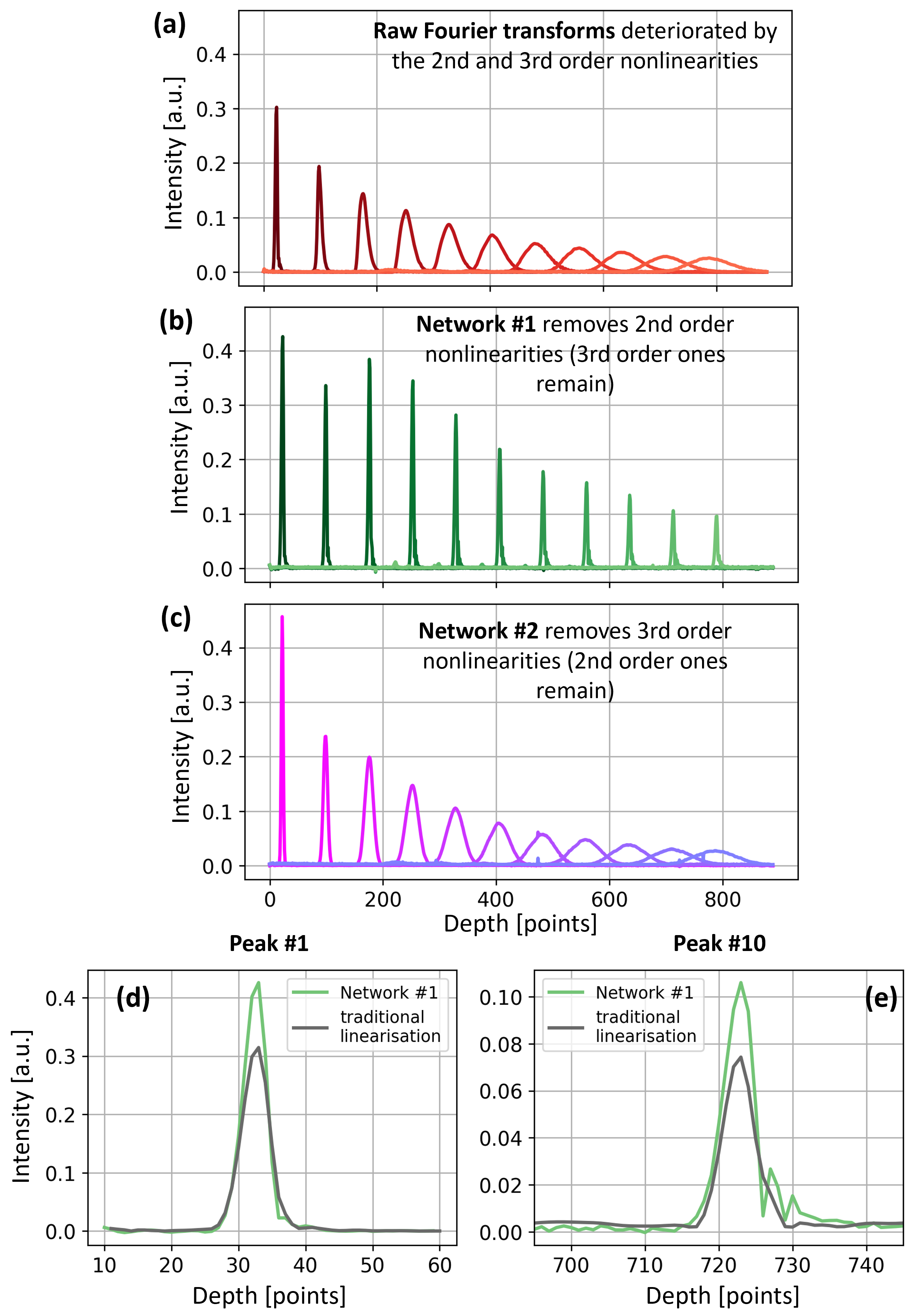}
    \captionsetup{belowskip=-6.5mm}
    \caption{Mirror at 11 different depths. (a) Fourier transforming raw signals results in 
  peaks that are both broadened by the second-order nonlinearity and distorted by the third-order nonlinearity. The deeper the peak, the bigger the nonlinearity effect, a behaviour characteristic of OCT signals acquired with a spectrometer. (b) Network~\textbf{\#1} removes the second-order nonlinearity, leaving the third-order one intact. (c) Network~\textbf{\#2} removes the third-order nonlinearity, leaving the third-order one intact. (d, e) The traditional linearisation (grey line) is not as precise as our approach (green lines). }
    \label{fig:mirrors}
\end{figure}

Networks were tested on the signals obtained through an OCT imaging of a mirror placed at 11 different depth positions. Each acquired signal Fourier transforms to a peak whose amount of second-order-nonlinearity-related broadening and third-order-nonlinearity-related distortion increases with depth (\cref{fig:mirrors}a). Due to their simultaneous appearance, the third-order nonlinearity manifests itself in the form of peak elongation. The depth-dependent deterioration is due to the fundamental limitations of OCT's spectrometer-based light detection. Additionally, the second- and third-order nonlinearities originating from the OCT's interferometer add up - depth-independently - to that effect, further worsening the outcome. Typically, spectrometer- and interferometer-related nonlinearities can be satisfactorily removed in a calibration step before the measurement.

The application of Network~\textbf{\#1} removes the second-order nonlinearity in its entirety and consequently, provides a dramatic reduction of peaks' widths (\cref{fig:mirrors}b). Because the third-order nonlinearity remains intact, the peaks' shapes incorporate the characteristic "tail". On the other hand, Network~\textbf{\#2} removes the third-order nonlinearity, leading to a complete reduction of the elongation (\cref{fig:mirrors}c).

 One observes the superiority of our approach when the peaks obtained with Network~\textbf{\#1} are compared to the peaks obtained as a result of traditional linearisation methods which remove all nonlinearity orders. Here, we used the method in \cite{wang2008spectral} where two raw signals corresponding to the mirror placed at two different locations are used. First, their phases are subtracted to remove the constant interferometer-related contribution, enabling the calculation of the correction vector for spectrometer-related nonlinearities. Once the spectrometer’s nonlinearity is removed, one of the initial spectra is used to generate an interferometer-nonlinearities-compensating vector. The peaks located close to 0 depth index obtained using our approach and the standard linearisation method (\cref{fig:mirrors}d, green and grey lines) are similar in shape: the network-outputted peak is higher which is very favourable - it means a higher signal-to-noise ratio, which is associated with better quality imaging. One observes a small "tail" at the side, most probably due to a small amount of third-order nonlinearity present at such shallow depths, whereas the peak obtained using the traditional methods is a smooth Gaussian. More interestingly, for a peak at a substantial depth (\cref{fig:mirrors}e), where the third-order nonlinearity is naturally much bigger (let us remind that this depth-increasing nonlinearity is tied to OCT spectrometer), the "tail" is much more pronounced in case of our network but not present in case of the traditional method. In the latter case, the peak is elongated which suggests the presence of both second- and third-order nonlinearities. This leads to the conclusion that traditional linearisation, although very satisfactory in practical applications, still has room for improvement. In fact, despite 30 years of OCT in science and industry, new methods for more precise OCT data linearisation are still being developed.

\subsection{Glass} \label{subsec:glass}

\begin{figure}[t]
    \centering
    \includegraphics[width=0.9\linewidth]{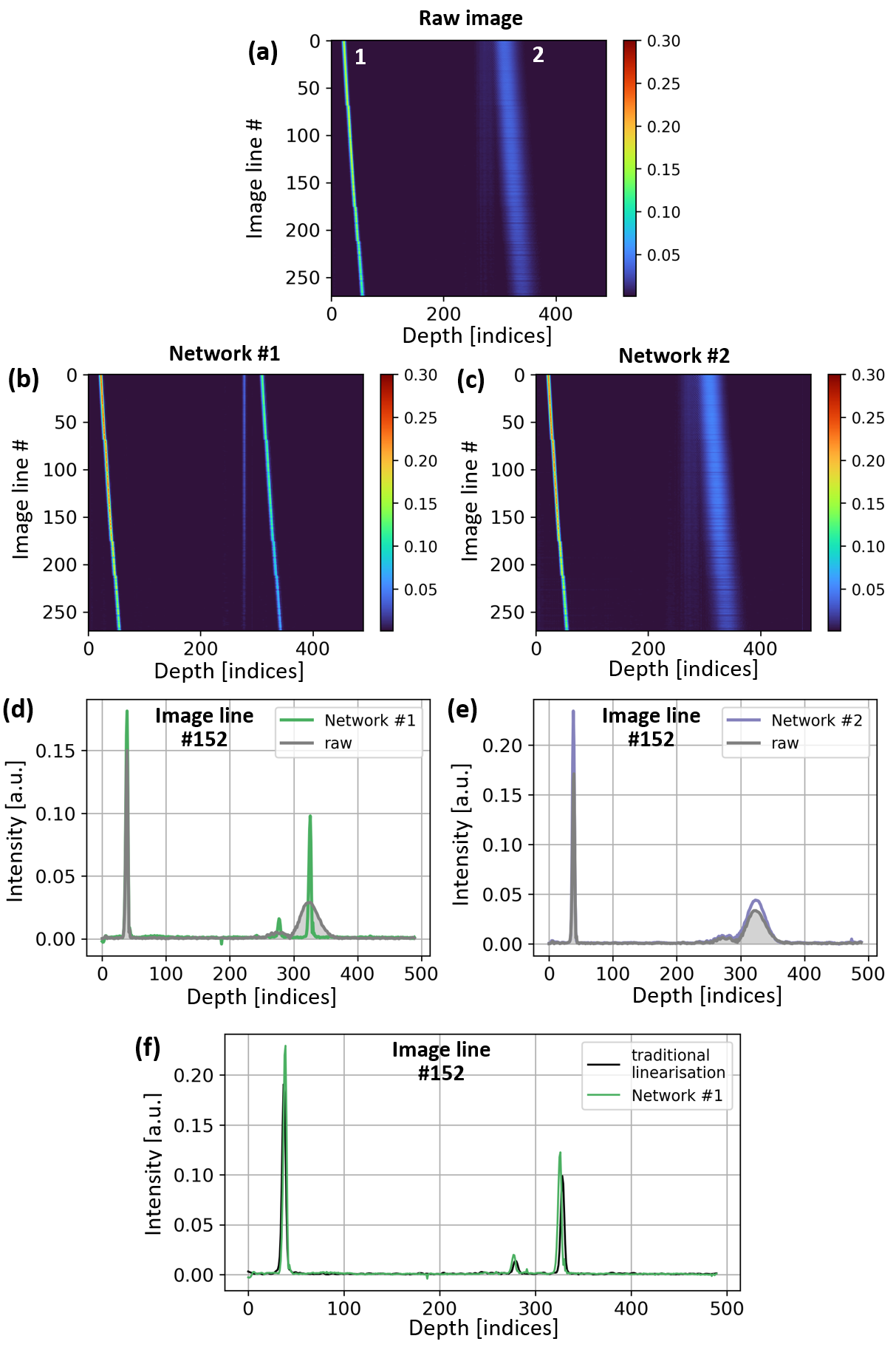}
    \captionsetup{belowskip=-6.5mm}
    \caption{Sapphire glass.
  (a) Fourier transformation of raw signals gives an OCT image where the quality deteriorates with depth. (b) Network~\textbf{\#1} removes the second-order nonlinearity which accounts for the most of quality deterioration, (c) Due to a small amount of the third-order nonlinearity, the application of Network~\textbf{\#2} does not improve the image quality. (d), Image line \#152 from Network~\textbf{\#1} (green line) is compared with its raw image equivalent (grey line) to show dramatic peaks' width reduction, (e) The image line \#152 from Network~\textbf{\#2} output (green line) and its raw image equivalent (grey line) show no significant difference except for a more favourable shape for the former. (f) Network~\textbf{\#1} (green line) performs quite well when compared to the traditional method output (black line). 1 and 2 mark the front and back surface of the imaged glass, the middle line is an artefact. }
    \label{fig:glass}
\end{figure}

Another test was performed on data corresponding to a 300$\mu$m thick sapphire glass put at an angle obtained with the same instrument. This data consisted of 256 raw OCT signals which represent a different transversal location on the glass. When Fourier transformed one after another, they form a 2D depth image of the object (\cref{fig:glass}a, 1 and 2 denote the front and back surface of the glass, the third line is an artefact placed at the depth equal to the thickness of the glass, inherent to OCT). Without linearisation, the image elements are broadened the deeper they are. When the same data is processed with Network~\textbf{\#1}, the broadening is removed from all the image elements (see the image in \cref{fig:glass}b and the 152nd image line in \cref{fig:glass}d). Due to very small amounts of the third-order nonlinearity, Network~\textbf{\#2} does not bring perceptible improvements (\cref{fig:glass}c), but under a closer inspection of a single image line (\cref{fig:glass}e), one can observe some improvement in the peak height and a possible side elongation reduction.

\cref{fig:glass}f compares the output of Network~\textbf{\#1} (green line) and the traditional linearisation method (black line), indicating that this model could be a satisfactory alternative for the traditional image calculation algorithm when the third-order nonlinearities are negligible. It can be used as such an alternative for the OCT laboratory instrument which was used here to obtain the data and which is nevertheless a fairly good representative of a standard OCT instrument. We note that Network~\textbf{\#1} improved the shape of the artefact (the middle peak in \cref{fig:glass}f), which - although generally irrelevant for OCT imaging - is still something noteworthy as such a feat is beyond the capabilities of traditional linearisation methods, showing that our approach, unlike the traditional ones, removes the nonlinearities nondiscriminatively.

\subsection{Grape and cucumber} \label{subsec:grapecucumber}

\begin{figure}[!t]
    \centering
    \includegraphics[width=0.9\linewidth]{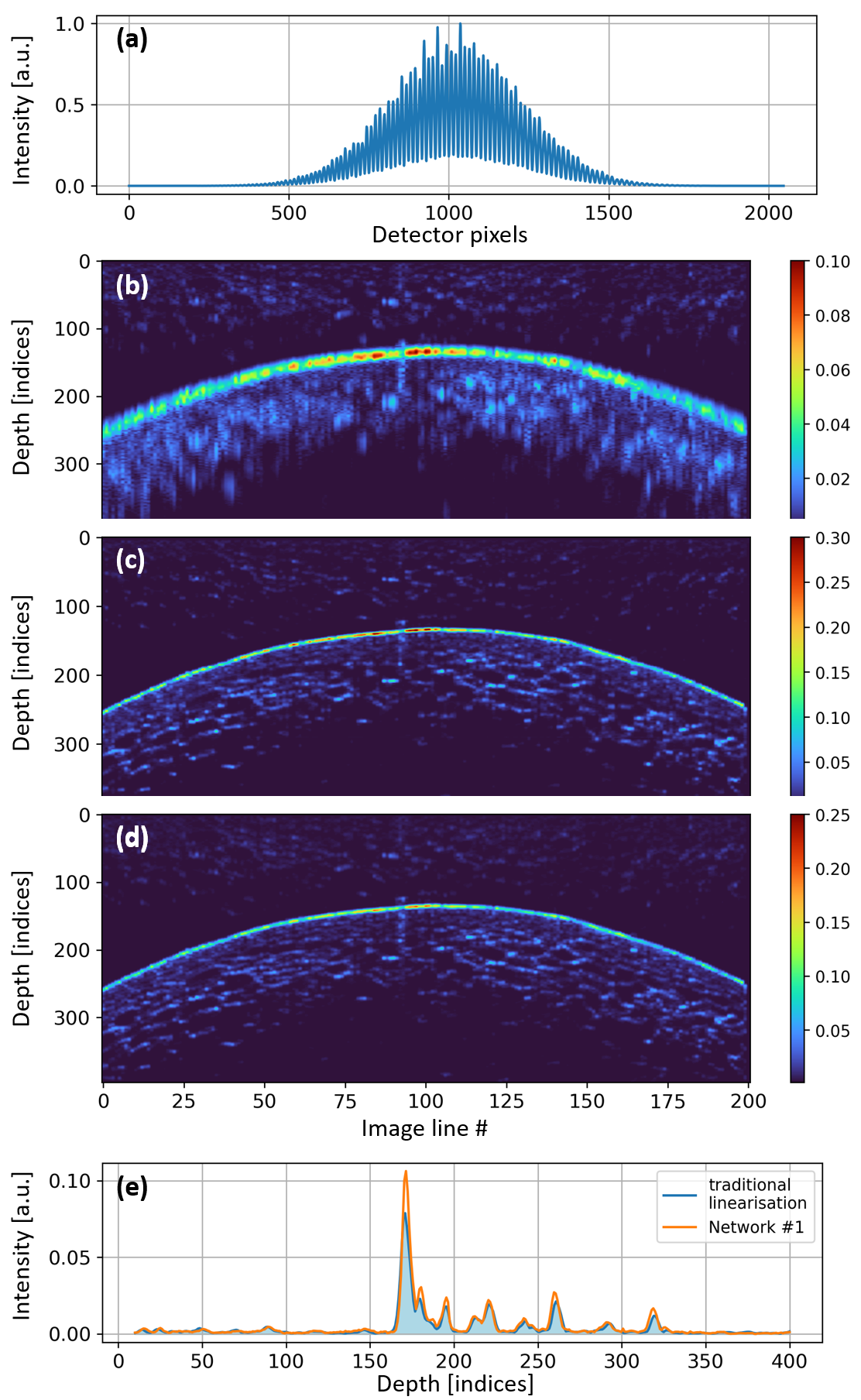}
    \caption{Grape.
  (a) One of the raw signals representing a single line in the image of a grape. (b) Fourier transform of raw signals gives a low-quality image. (c) Network~\textbf{\#1} removes the second-order nonlinearity providing a high-quality image when compared with (d) the image obtained using traditional linearisation methods. (e) The prediction for image line \#200 agrees very well with its equivalent from the image obtained using traditional methods. }
    \label{fig:grape}
    \vspace{-5mm}
\end{figure}

\begin{figure}[!t]
    \centering
    \includegraphics[width=0.9\linewidth]{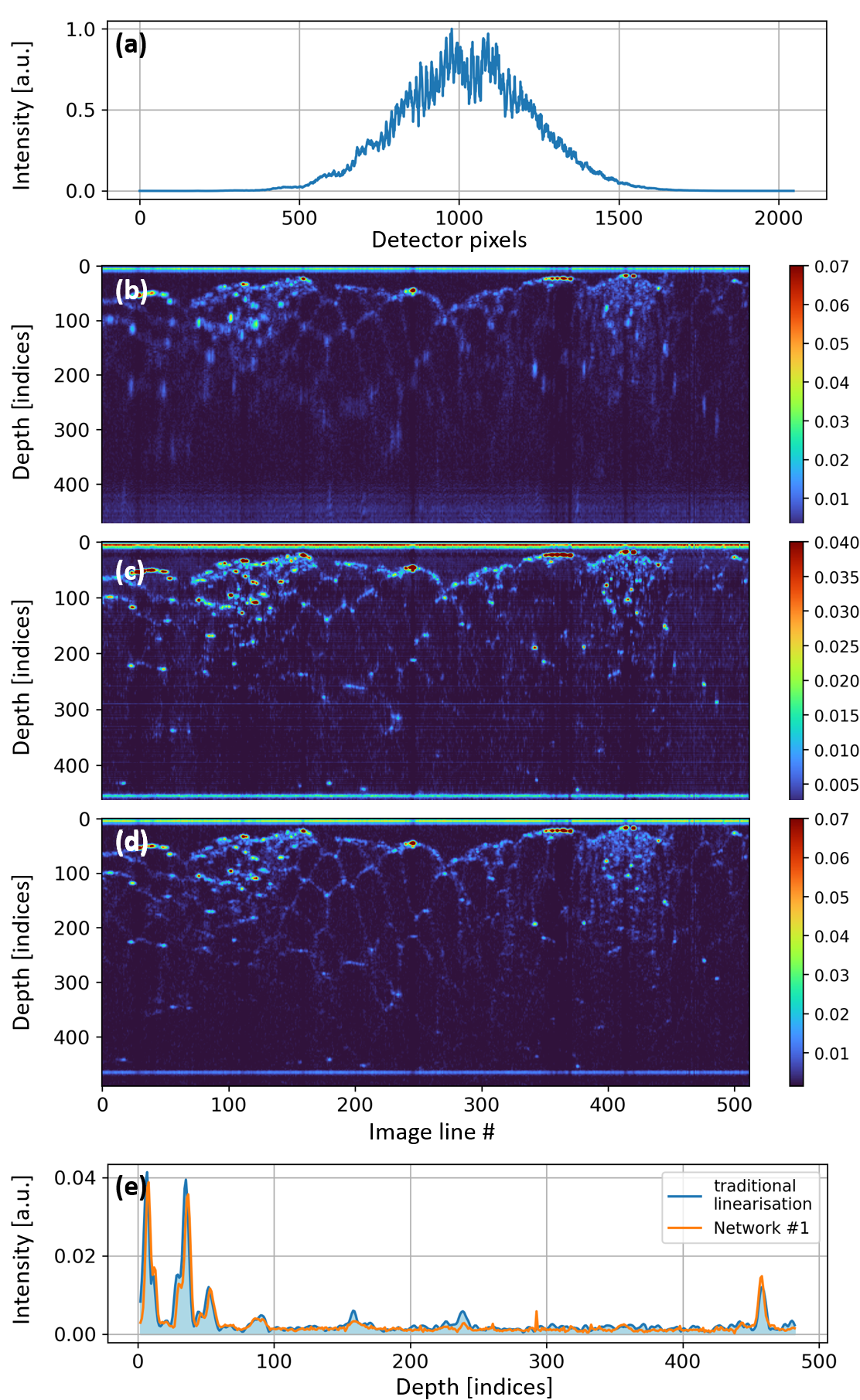}
    \caption{Cucumber.
  (a) One of the raw signals representing a single line in the image of a cucumber. (b) Fourier transform of raw signals gives a low-quality image. (c) Network~\textbf{\#1} removes the second-order nonlinearity providing a high-quality image compared with (d) the image obtained using traditional linearisation methods. (e) The prediction for image line \#200 agrees well with its equivalent from the image obtained using traditional methods. }
    \label{fig:cucumber}
    \vspace{-5mm}
\end{figure}

The final tests were performed on the data corresponding to a grape (\cref{fig:grape}) and a cucumber (\cref{fig:cucumber}). The former represents a high signal-to-noise ratio (highly visible modulation in the signal as depicted in \cref{fig:grape}a) and the latter a low signal-to-noise ratio (poorly visible modulation in the signal as depicted in \cref{fig:cucumber}a).

Fourier transformation of raw signals gives images of a low quality that further deteriorates with depth (\cref{fig:grape}b, \cref{fig:cucumber}b). Again, the improvement obtained when applying Network~\textbf{\#1} (\cref{fig:grape}c, \cref{fig:cucumber}c) is comparable to that obtained with traditional methods (\cref{fig:grape}d, \cref{fig:cucumber}d). Under closer inspection (\cref{fig:grape}e, \cref{fig:cucumber}e), one notices that Network~\textbf{\#1} output is less detailed for the higher signal-to-noise grape data - most probably due to the remaining third-order nonlinearities - and loses some details for the lower signal-to-noise ratio cucumber data. The latter loss is the direct consequence of training data representing predominantly high signal-to-noise ratio signals. 

Just as in the case of the glass data, due to small amounts of third-order nonlinearity, Network~\textbf{\#2} did not provide visually significant improvement and therefore, was omitted.

\section{Summary and discussion} \label{sec:summary}

Two models were demonstrated: one for removing second-order nonlinearity from digital signals and leaving the third-order one intact, and another one for removing the third-order nonlinearity but leaving the second-order one intact. Their nonlinearity-removing capabilities were extensively analysed using computer-generated signals as well as experimental signals acquired with a laboratory Optical Coherence Tomography (OCT) device. Tests on the OCT signals, which corresponded to objects with varying levels of complication: mirror, sapphire, grape and cucumber, allowed us to evaluate the networks in terms of their applicability in a world-relevant technology. Whereas the tests and analysis done on the computer-generated signals revealed extraordinary performance in terms of similarity of the ground truth and prediction signals, the tests on OCT-acquired data indicated the networks' advantageous operation when compared to approaches traditionally used in OCT.

During training, both networks converged exceptionally fast and reached remarkable loss function values. Nevertheless, we have identified 
possible improvements: 
the closer inspection of the predictions revealed non-zero values that are zero-valued areas in the ground-truth signals. Another current shortcoming of both networks is their time of processing an entire OCT image; consequently, for now, the processing speed is insufficient for the real-time processing of data coming straight from OCT devices. 

\section{Future work} \label{sec:future}

The future work will mainly consist in developing an approach that enables one to use both networks in a sequence so that both the second- and third-order nonlinearity removal is achieved. Since, as was shown here, the networks remove the nonlinearities completely, their sequential use will provide a very valuable tool for extremely precise data linearisation, especially attractive for OCT where any uncompensated nonlinearities lead to the quality drop.

Also, in terms of the network performance, we will try to remove the discrepancy for zero-valued elements in the ground truth and prediction signals by adding extra convolutional layers before the output to filter out the unwanted values as well as performing additional hyperparameter tuning. Additionally, we plan to speed up the processing of the stacks by lowering the amount of calculations and the size of the networks.

\noindent\textbf{Acknowledgements.} Supported by the New Zealand Ministry of Business, Innovation and Employment (MBIE), Smart Ideas grant “Extending the Boundaries of Digital Signal Processing: AI-powered Fourier Transformation Alternative (E7943)”. We also thank Oliver Batchelor, François Bissey, and Piotr Kolenderski for sharing their computational resources.
\newpage
\clearpage
{
    \small
    \bibliographystyle{unsrtnat}
    \bibliography{main}
}


\end{document}